\documentclass[aps,prd,groupaddress,floatfix,tighten,nofootinbib,showpacs,amsfonts,superscriptaddress]{revtex4-2}
\usepackage{url}
\usepackage{xcolor}
\usepackage{hyperref}
\usepackage[latin1]{inputenc}

\colorlet{shadecolor}{gray!15}
\definecolor{greenLinks}{rgb}{0, 0.6, 0}
\definecolor{blueLinks}{rgb}{0, 0, 0.6}
\definecolor{redLinks}{rgb}{0.6, 0, 0}
\definecolor{tempText}{rgb}{0.55, 0.10,0.67}
\definecolor{eprintLinks}{rgb}{0.4, 0.4, 0.4}
\definecolor{journalLinks}{rgb}{0.6, 0, 0}
\usepackage{amssymb}
\usepackage{amsmath,color}
\usepackage{epsfig}
\usepackage{graphicx,epsf,epsfig}
\usepackage{footnote}
\usepackage{multirow}
\usepackage{enumitem}
\usepackage{color}
\def\slc#1{\setbox0=\hbox{$#1$}                  % set a box for #1
    \dimen0=\wd0                                 % and get its size
    \setbox1=\hbox{/} \dimen1=\wd1               % get size of /
    \ifdim\dimen0>\dimen1                        % #1 is bigger
       \rlap{\hbox to \dimen0{\hfil/\hfil}}      % so center / in box
       #1                                        % and print #1
    \else                                        % / is bigger
       \rlap{\hbox to \dimen1{\hfil$#1$\hfil}}   % so center #1
       /                                         % and print /
    \fi}

\def\be{\begin{equation}}
\def\ee{\end{equation}}
\def\gs{\mathrel{
   \rlap{\raise 0.511ex \hbox{$>$}}{\lower 0.511ex \hbox{$\sim$}}}}
\def\ls{\mathrel{
   \rlap{\raise 0.511ex \hbox{$<$}}{\lower 0.511ex \hbox{$\sim$}}}}

\newcommand{\ba}{\begin{array}{c}}
\newcommand{\baz}{\begin{array}{cc}}
\newcommand{\barrr}{\begin{array}{rrr}}
\newcommand{\bad}{\begin{array}{ccc}}
\newcommand{\bav}{\begin{array}{cccc}}
\newcommand{\baf}{\begin{array}{ccccc}}
\newcommand{\bea}{\begin{equation} \begin{array}{c}}
\newcommand{\eea}{\end{array} \end{equation}}
\newcommand{\ea}{\end{array}}

\usepackage[normalem]{ulem}

\def\21{$\mathrm{SU(2)_L \otimes U(1)_Y}$ }

\newcommand {\ignore}[1]{}

\newcommand{\nn}{\nonumber}

\allowdisplaybreaks \allowdisplaybreaks[2]
\newcommand{\AddrCECYT}{Centro de Estudios Cient\'ificos y Tecnol\'ogicos No 16, Instituto Polit\'ecnico Nacional, Pachuca: Ciudad del Conocimiento y la Cultura, Carretera Pachuca Actopan km 1+500, San Agust\'in Tlaxiaca, Hidalgo, M\'exico.\\ 
}

\begin{document}
%-----------------------------------------------------------------------------
\title{A non-renormalizable neutrino mass model with $\mathbf{S}_{3}\otimes \mathbf{Z}_{2}$ symmetry} 
%-----------------------------------------------------------------------------
%
\author{J. D. Garc\'ia-Aguilar}
\email{jdgarcia@ipn.mx}
\affiliation{\AddrCECYT}

\author{Juan Carlos G\'omez-Izquierdo}
\email{cizquierdo@ipn.mx\\
}
\affiliation{\AddrCECYT}
%

%

%
%-----------------------------------------------------------------------------
%\pacs{14.60.Pq, 11.30.Er}

\date{\bf \today} 

\begin{abstract}\vspace{2cm}

The lepton sector is studied within a flavored non-renormalizable model where 
the $\mathbf{S}_{3}\otimes \mathbf{Z}_{2}$ flavor symmetry drives the Yukawa couplings. In this framework, the effective neutrino mass, that comes from the type II see-saw mechanism, as well as  the charged lepton mass matrices are hierarchical and these have (under a benchmark in the charged sector) a kind of Fritzsch textures that accommodate the mixing angles in good agreement with the last experimental data. The model favors the normal hierarchy, this also predicts  consistent values for the CP-violating phase and the $\vert m_{ee}\vert$ effective Majorana neutrino mass rate. Along with this, the branching ratio for the lepton flavor violation process, $\mu\rightarrow e\gamma$, is well below the current bound.

%a CP-violating phase consistent with the experimental data and the branching ratio for the lepton flavor violation process, $\mu\rightarrow e\gamma$, is well below the current bound.

\end{abstract}

\begin{flushright}
CECyT 16-21-I
\end{flushright}

%-----------------------------------------------------------------------------
\maketitle
%-----------------------------------------------------------------------------
%

\section{Introduction}

In spite of the fact that Standard Model (SM) works
almost perfectly, it fails to explain the neutrino experimental data, dark matter, baryon asymmetry of the universe and so forth \cite{Masiero:2005ua}. Speaking about the mixings, the lepton sector exhibits a peculiar pattern which is totally different to the quark sector where
the mixing matrix is almost diagonal and this puzzle remains unsolved.

In this line of thought, hierarchical quark mass matrices
as the nearest neighbor interaction (NNI) textures~\cite{Branco:1988iq, Branco:1994jx, Harayama:1996am, Harayama:1996jr} and those that 
possess the generalized Fritzsch textures ~\cite{Fritzsch:2015gxa}, fit quite well the CKM  matrix ~\cite{Cabibbo:1963yz, Kobayashi:1973fv}. 
In the lepton sector, according to the experimental data,
the PMNS matrix ~\cite{Maki:1962mu, Pontecorvo:1967fh} has large values in its entries which can be understood by the presence of a symmetry behind the neutrino mass matrix. Currently, we can find in the literature elegant proposals (and their respective breaking) as the $\mu\leftrightarrow \tau$ symmetry~\cite{Fukuyama:1997ky, Mohapatra:1998ka, Lam:2001fb, Kitabayashi:2002jd, Grimus:2003kq, Xing:2015fdg,Koide:2003rx, Fukuyama:2020swd},
$\mu\leftrightarrow \tau$ reflection symmetry \cite{Ahn:2008hy, Nishi:2016wki, Chen:2016ica, Chen:2015siy, Zhao:2017yvw, Liu:2017frs, Zhao:2018vxy,Nath:2018hjx}, Tri-Bimaximal~\cite{Harrison2002167, Xing200285, Altarelli:2012ss, Rahat:2018sgs, Perez:2019aqq, Rahat:2020mio}, Cobimaximal mixing matrices \cite{Fukuura:1999ze, Miura:2000sx, Ma:2002ce,Ferreira:2016sbb, Ma:2016nkf, Ma:2017moj, Ma:2017trv, Grimus:2017itg,CarcamoHernandez:2017owh,CarcamoHernandez:2018hst,CarcamoHernandez:2020udg}. Moreover, hierarchical mass matrices as the Fritzsch~ \cite{Fritzsch:2015foa} and the generalized Fritzsch textures~\cite{Fritzsch:2015gxa} also accommodate quite well the PMNS mixing matrix.

From the model building point of view, the flavor symmetries~\cite{Ishimori:2010au,Grimus:2011fk,Ishimori:2012zz,King:2013eh} have been useful to get desirable textures in the fermion mass matrices, and therefore, the well known mixing patterns. For example, the $\mathbf{S}_{3}$ non-abelian group that has been explored exhaustively in different frameworks~\cite{Pakvasa:1977in, Gerard:1982mm, Kubo:2003iw, Kubo:2003pd,Kobayashi:2003fh, Chen:2004rr, Kubo:2005sr, Felix:2006pn, Mondragon:2007af, Mondragon:2007nk, Mondragon:2007jx, Meloni:2010aw, Dicus:2010iq, Dong:2011vb, Canales:2011ug, Canales:2012ix, Kubo:2012ty, Canales:2012dr, GonzalezCanales:2012kj, Dias:2012bh, GonzalezCanales:2012za, Meloni:2012ci, Canales:2013ura, Ma:2013zca, Canales:2013cga, Hernandez:2014lpa, Hernandez:2014vta, Ma:2014qra, Gupta:2014nba, Hernandez:2015dga, Hernandez:2015zeh, Hernandez:2015hrt, Arbelaez:2016mhg, Hernandez:2013hea, CarcamoHernandez:2016pdu, Das:2014fea, Das:2015sca, Pramanick:2016mdp, Das:2017zrm, Cruz:2017add, Gomez-Izquierdo:2017rxi, Garces:2018nar, Gomez-Izquierdo:2018jrx, Ge:2018ofp, Das:2018rdf, Xing:2019edp, Pramanick:2019oxb, Kuncinas:2020wrn, Vien:2020trr, Espinoza:2018itz,Espinoza:2020qyf}. In the mentioned literature there are few models \cite{Meloni:2010aw, Dias:2012bh} where the Fritzsch textures have been implemented. Hence, the main purpose that we pursuit is to realize those textures by means the $\mathbf{S}_{3}$ flavor symmetry, however, we obtain a modified Fritzsch textures which are different to previous studies.

Due to the last neutrino oscillations data seem to favor the normal hierarchy \cite{deSalas:2020pgw}, in this paper, we construct a non-renormalizable lepton model in the type II see-saw scenario where the $\mathbf{S}_{3}\otimes \mathbf{Z}_{2}$ flavor symmetry  drives the Yukawa couplings. We stress that the scalar sector of the mentioned model keeps intact so that flavons are included to generate the mixings. In this work, the effective neutrino as well as  the charged lepton mass matrices are hierarchical and these have (under a benchmark in the charged sector) a kind of Fritzsch textures that accommodate the mixing angles in good agreement with the last experimental data. The model predicts  consistent values for the  CP-violating phase and the $\vert m_{ee}\vert$ effective Majorana neutrino mass rate. Along with this, the branching ratio for the lepton flavor violation process, $\mu\rightarrow e\gamma$, is well below the current bound.

The plan of the paper is as follows: the framework, the matter content of the model and the fermion mass matrices are described in detail in section II; in the section III, the PMNS mixing matrix is obtained and relevant features are remarked. An analytical study is carried out on the mixing angles to find the parameter space that accommodates the observables, this together with a numerical study in section IV. In section V, we give some model predictions and relevant conclusions are shown in section VI.

\section{The framework}

The current framework is a scalar extension of the SM so that the usual matter content under the gauge group $\mathbf{SU(3)}_{C}\otimes \mathbf{SU(2)}_{L} \otimes \mathbf{U(1)}_{Y}$ is considered. Explicitly, the fields are
\begin{eqnarray}
Q_{L}&=&\begin{pmatrix}
u_{L}\\ d_{L}
\end{pmatrix}\sim \left(3, 2, \frac{1}{3}\right), \qquad d_{R}\sim \left(3, 1,-\frac{2}{3}\right),\qquad u_{R}\sim \left(3, 1,\frac{4}{3}\right);\nn\\
L&=&\begin{pmatrix}
\nu_{L}\\ e_{L}
\end{pmatrix}\sim \left(1, 2,-1\right),\qquad e_{R}\sim \left(1, 1,-2\right).
\end{eqnarray}
Additionally, in the scalar sector we have the following fields
\begin{equation}\label{scal}
H=\begin{pmatrix}
H^{+}\\ H^{0}
\end{pmatrix}\sim \left(1, 2, 1\right),\quad \Delta= \left( 
\begin{array}{cc}
\frac{\Delta ^{+}}{\sqrt{2}} & \Delta ^{++} \\ 
\Delta^{0} & -\frac{\Delta ^{+}}{\sqrt{2}}%
\end{array}%
\right)\sim \left(1, 3, 2\right).
\end{equation}

Having introduced the matter content, the Yukawa mass term is given by
\begin{equation}
\mathcal{L}=\mathcal{L}_{SM}-\frac{1}{2}Y^{\nu}\bar{L}(i\sigma_{2})\Delta \left(L\right)^{c}-V(H,\Delta)+h.c.
\end{equation}

Although the quark and scalar fields have been mentioned, the quark mixings and the scalar potential analysis will leave out in this work. We have to point out that scalar potential analysis is crucial to get a viable model but the full study is a working progress.

Speaking about the flavor symmetry, we will use the $\mathbf{S}_{3}$~\cite{Ishimori:2010au} due to the
three dimensional real representation can be decomposed as: ${\bf 3}_{S}={\bf 2}\oplus {\bf 1}_{S}$ or ${\bf 3}_{A}={\bf 2}\oplus {\bf 1}_{A}$. This structure seems to work quite well for obtaining hierarchical mass matrices. Along with this, the $\mathbf{Z}_{2}$ discrete symmetry can be used to forbid some Yukawa couplings, in our work this is needed to prohibit the renormalizable terms.

\subsection{The model}
As we already commented, in the present model, the scalar sector contains one Higgs doublet ($H$) and one triplet ($\Delta$) so that some flavons
will be added to the matter content in order to generate the mass textures that provide the mixings. Then, the matter fields transform in a non trivial way. Hence, the assignation under the $\mathbf{S}_{3}\otimes \mathbf{Z}_{2}$ is shown in the following table.
\begin{table}[ht]
	\begin{center}
		\begin{tabular}{|c|c|c|c|c|c|c|c|c|c|c|}
			\hline\hline
			{\footnotesize Matter}	& {\footnotesize $L_{I}$} & {\footnotesize $L_{3}$} & {\footnotesize $e_{I R}$}  & {\footnotesize $e_{3 R}$} & {\footnotesize $\phi_{I}$} & {\footnotesize $\phi_{3}$} & {\footnotesize $\Delta$} & {\footnotesize $H$} \\
			\hline
			{\footnotesize $\mathbf{S}_{3}$}	& {\footnotesize \bf $2$} & {\footnotesize \bf $1_{S}$} & {\footnotesize \bf $2$} & {\footnotesize \bf $1_{S}$} & {\footnotesize \bf $2$} & {\footnotesize \bf $1_{S}$} & {\footnotesize \bf $1_{S}$} & {\footnotesize \bf $1_{S}$}\\
			\hline
			{\footnotesize  $\mathbf{Z}_{2}$}	& 1 & 1 & 1 & 1 & -1 & -1 & -1 & -1 \\\hline\hline
		\end{tabular}\caption{Matter content for the lepton sector. $I=1,2$}
	\end{center}
\end{table}

As one can notice, due to the $\mathbf{Z}_{2}$ symmetry there are no-renormalizable Yukawa mass term, then, at the next leading order in the cutoff scale we have
%According to the matter content and its respective assignation under $\mathbf{S}_{3}\otimes \mathbf{Z}_{2}$ symmetry there are no renormalizable Yukawa mass terms but at the next leading order in the cutoff scale, the allowed Yukawa mass term is given as
\begin{eqnarray}
-\mathcal{L}_{Y}&=&
\frac{y_{1}^{e}}{\Lambda}\left[ \bar{L}_{1}H\left(
\phi_{1}e_{2R}+\phi_{2}e_{1R}\right) +\bar{L}_{2}H\left(
\phi_{1}e_{1R}-\phi_{2}e_{2R}\right) \right] +\frac{y_{2}^{e}}{\Lambda}\left[ \bar{L}%
_{1}H \phi_{3}e_{1R}+\bar{L}_{2}H\phi_{3}e_{2R}\right] +\frac{y_{3}^{e}}{\Lambda}\left[ \bar{L}%
_{1}H \phi_{1}+\bar{L}_{2}H \phi_{2}\right]e_{3R} \notag \\
&&+\frac{y_{4}^{e}}{\Lambda}\bar{L}_{3}\left[ H \phi_{1}e_{1R}+H \phi_{2}e_{2R}\right] +\frac{y_{5}^{e}}{\Lambda}\bar{L}_{3}H \phi_{3}e_{3R}
+\frac{y_{1}^{\nu}}{\Lambda}\left[ \bar{L}_{1}\Delta\left(
\phi_{1}L_{2}+\phi_{2}L_{1}\right) +\bar{L}_{2}\Delta\left(
\phi_{1}L_{1}-\phi_{2}L_{2}\right) \right]
\notag \\&&+
\frac{y_{2}^{\nu}}{\Lambda}\left[ \bar{L}%
_{1}\Delta \phi_{3}L_{1}+\bar{L}_{2}\Delta\phi_{3}L_{2}\right] +\frac{y_{3}^{\nu}}{\Lambda}\left[ \bar{L}%
_{1}\Delta \phi_{1}+\bar{L}_{2}\Delta \phi_{2}\right] L_{3}+\frac{y_{4}^{\nu}}{\Lambda}\bar{L}_{3}\left[ \Delta \phi_{1}L_{1}+\Delta \phi_{2}L_{2}\right] +\frac{y_{5}^{\nu}}{\Lambda}\bar{L}_{3}\Delta \phi_{3} L_{3}
+h.c.
\end{eqnarray}
As result of this, the lepton mass matrices are given as
\begin{equation}
{\bf M}_{e}=\begin{pmatrix}
a_{e}+b^{\prime}_{e} & b_{e} & c_{e} \\ 
b_{e} & a_{e}-b^{\prime}_{e} & c^{\prime}_{e} \\ 
f_{e} & f^{\prime}_{e} & g_{e}
\end{pmatrix},\qquad {\bf M}_{\nu}=\begin{pmatrix}
a_{\nu}+b^{\prime}_{\nu} & b_{\nu} & c_{\nu} \\ 
b_{\nu} & a_{\nu}-b^{\prime}_{\nu} & c^{\prime}_{\nu} \\ 
c_{\nu} & c^{\prime}_{\nu} & g_{\nu}
\end{pmatrix}
\end{equation}
with 
\begin{eqnarray}
a_{e} &=&y_{2}^{e}v \frac{\langle \phi_{3}\rangle}{\Lambda},\quad b_{e}^{\prime }=y_{1}^{e} v\frac{\langle \phi_{2}\rangle}{\Lambda},\quad
b_{e}=y_{1}^{e}v\frac{\langle \phi_{1}\rangle}{\Lambda},\quad c_{e}=y_{3}^{e}v\frac{\langle \phi_{1}\rangle}{\Lambda};\notag \\ c_{e}^{\prime
}&=&y_{3}^{e}v\frac{\langle \phi_{2}\rangle}{\Lambda},\quad f_{e}=y_{4}^{e}v\frac{\langle \phi_{1}\rangle}{\Lambda},\quad
f_{e}^{\prime } =y_{4}^{e} v\frac{\langle \phi_{2}\rangle}{\Lambda},\quad g_{e}=y_{5}^{e} v\frac{\langle \phi_{3}\rangle}{\Lambda};\notag \\
a_{\nu} &=&y_{2}^{\nu}v_{\Delta} \frac{\langle \phi_{3}\rangle}{\Lambda} ,\quad b_{\nu}^{\prime }=y_{1}^{\nu} v_{\Delta} \frac{\langle \phi_{2}\rangle}{\Lambda},\quad
b_{\nu}=y_{1}^{\nu}v_{\Delta} \frac{\langle \phi_{1}\rangle}{\Lambda};\notag \\ c_{\nu}&=&y_{3}^{\nu}v_{\Delta} \frac{\langle \phi_{1}\rangle}{\Lambda},\quad c_{\nu}^{\prime
}=y_{3}^{\nu}v_{\Delta} \frac{\langle \phi_{2}\rangle}{\Lambda},\quad g_{\nu}=y_{5}^{\nu}v_{\Delta} \frac{\langle \phi_{3}\rangle}{\Lambda}.
\end{eqnarray}

Here, $v$ and $v_{\Delta}$ stand for the vacuum expectation values (vev's) of the Higgs doublet and triplet, respectively. In order to reduce the free parameters in the lepton mass matrices, we assume the following vev's pattern for the flavon doublet and singlet of $\mathbf{S}_{3}$, respectively: $\langle \phi\rangle=v_{\phi}(1, 0)$ and $\langle \phi_{3}\rangle=v_{\phi_{3}}$ \footnote{In fact, one might consider two different vev's alignments: (a)  $\langle \phi\rangle=v_{\phi}(0, 1)$ and $\langle \phi_{3}\rangle=v_{\phi_{3}}$ but this does not provide the right mixings; (b) $\langle \phi\rangle=v_{\phi}(1, 1)$ and $\langle \phi_{3}\rangle=v_{3}$, in this case, the free parameters increase.}. At the same time, we set the magnitudes of the vev's as follows: $v_{\phi}\sim \lambda \Lambda$ and $v_{\phi_{3}}\sim\lambda \Lambda$ where $\lambda=0.225$ is the Wolfenstein parameter. Before finishing this section, we would like to remark that the flavor symmetry is broken by the vev's of the flavons and the cutoff $\Lambda$ scale satisfies the hierarchy $\Lambda\gg v\gg v_{\Delta}$. Therefore, the main role that the flavons play is to provide the mixings as was already commented.

%We have to keep in mind that $\langle \phi_{i} \rangle/\Lambda\sim \lambda=0.225$ where the Wolfenstein parameter ($\lambda$) is a typical factor to suppress the matrix elements in non-renormalizeble models.

%Before finishing this section, we would like to remark that the flavor symmetry is broken by the vacuum expectation values (vev's) of the flavons and the cutoff $\Lambda$ scale is larger than the electroweak and the left ones. The main role that plays the flavons is provided the mixings as was already commented.

%A new mechanism to symmetry breaking is studied taking into account a five dimensional spacetime where Lorentz invariance is violated in an explicit way. For that, we present a toy model with a global SU(2) symmetry whose four dimensional effective model recovers the Lorentz symmetry whereas the internal symmetry is broken in the Kaluza-Klein sector. Moreover, we show that radiative corrections will communicate the symmetry breaking at the zero mode level.

%We propose a new mechanism to break the symmetry, this consists in taking into account a five dimensional spacetime where the Lorentz invariance is violated explicitly. To do this,  we present a toy model  with a global SU(2) symmetry whose four dimensional effective model recovers the Lorentz symmetry whereas the internal symmetry is broken in the Kaluza-Klein sector. Moreover, we show that radiative corrections will communicate to the symmetry breaking at the zero mode level.

\section{PMNS MIXING MATRIX}

Due to the alignment, the mass matrices read as
%In order to reduce the number of free parameters,
%let us consider the following alignment for vev's of  the flavons $\langle \phi_{I}\rangle=(\langle \phi_{1} \rangle, 0 )$. Then, the masses read as

\begin{equation}
{\bf M}_{e}=\begin{pmatrix}
a_{e} & b_{e} & c_{e} \\ 
b_{e} & a_{e} & 0 \\ 
f_{e} & 0 & g_{e}
\end{pmatrix},\qquad {\bf M}_{\nu}=\begin{pmatrix}
a_{\nu} & b_{\nu} & c_{\nu} \\ 
b_{\nu} & a_{\nu} & 0 \\ 
c_{\nu} & 0 & g_{\nu}
\end{pmatrix}. 
\end{equation}

As one can notice, if $a_{e}$ ($a_{\nu}$) was zero, the charged lepton (neutrino) mass matrix  would possess implicitly the NNI (Fritzsch \footnote{ As is well known, the Fritzsch textures are given by
\begin{equation}
\mathbf{M}=\begin{pmatrix}
0 & A & 0 \\ 
A^{\ast} & 0 & B \\ 
0 & B^{\ast} & C
\end{pmatrix}.
\end{equation}
}) textures. In general, the charged lepton mass matrix has five complex free parameters, then, in order to reduce a little bit more the free parameters we will adopt the benchmark $c_{e}\approx f_{e}$. As a result, the lepton mass matrices have the Fritzsch textures but the entry $a_{(\nu, e)}$ will modify slightly those textures, as we will show next.

%in order to reduce a little bit more the free parameters in the charged lepton mass matrix, we will adopt the benchmark $c_{e}\approx f_{e}$. In this way, the lepton mass matrices have the Fritzsch textures but displacement them by the term $a_{(\nu, e)}$, as we will show next.

The mixing matrices that take place in the PMNS matrix are obtained as follows: $\mathbf{M}_{e}$ and $\mathbf{M}_{\nu}$ are diagonalized respectively by $\mathbf{U}_{e(L,R)}$ and $\mathbf{U}_{\nu}$ such that  
${\bf U}^{\dagger}_{e L} {\bf M}_{e}
{\bf U}_{e R}=\hat{\bf M}_{e}$ and  ${\bf U}^{\dagger}_{\nu} {\bf M}_{\nu}
{\bf U}^{\ast}_{\nu}=\hat{\bf M}_{\nu}$ with $\mathbf{\hat{M}}_{(e, \nu)}=\textrm{Diag.}(m_{(e,1)}, m_{(\mu, 2)}, m_{(\tau,3)})$ being the physical lepton masses. Then, we make the following rotation ${\bf U}_{e(L, R)}={\bf S}_{12} {\bf u}_{e(L,R)}$ and
${\bf U}_{\nu}={\bf S}_{12} {\bf u}_{\nu}$ 
so that one obtains $\mathbf{u}^{\dagger}_{e L} \mathbf{m}_{e}\mathbf{u}_{e R}=\mathbf{\hat{M}}_{e}$ 
and $\mathbf{u}^{\dagger}_{\nu} \mathbf{m}_{\nu}\mathbf{u}^{\ast}_{\nu }=\mathbf{\hat{M}}_{\nu}$  where  $\mathbf{m}_{(e,\nu)}$ and $\mathbf{S}_{12}$ are given respectively as

\begin{equation}
{\bf m}_{\ell}=\begin{pmatrix}
a_{\ell} & b_{\ell} & 0 \\ 
b_{\ell} & a_{\ell} & c_{\ell} \\ 
0 & c_{\ell} & g_{\ell}
\end{pmatrix},\qquad \mathbf{S}_{12}= \begin{pmatrix}
0 & 1 & 0 \\ 
1 & 0 & 0 \\ 
0 & 0 & 1
\end{pmatrix},
\end{equation}
where $\ell=\nu, e$. 

We can observe that both mass matrices can be written as
\begin{equation}
{\bf m}_{\ell}=a_{\ell}\mathbf{1}_{3\times 3}+\begin{pmatrix}
0 & b_{\ell} & 0 \\ 
b_{\ell} & 0& c_{\ell} \\ 
0 & c_{\ell} & g_{\ell}-a_{\ell}
\end{pmatrix}.
\end{equation}

As one can realize, the second mass matrix has the Fritzsch texture but the there is a shift due to the $a_{\ell}$ parameter. Consequently, we expect a deviation to the Fritzsch prediction on the mixings. Let us diagonalize the mass matrix, ${\bf m}_{\ell}$, where the CP violating phases are factorized as $\mathbf{m}_{\ell}=\mathbf{P}_{\ell}\bar{\mathbf{m}}_{\ell}\mathbf{P_{\ell}}$. Explicitly, we obtain

%The $a_{\ell}$ parameter will modify slightly the Fritzsch texture as we will see in the diagonalization procedure. First at all, the CP violating phases are factorized as $\mathbf{m}_{\ell}=\mathbf{P}_{\ell}\bar{\mathbf{m}}_{\ell}\mathbf{P_{\ell}}$. Explicitly, we obtain

\begin{eqnarray}
\mathbf{P}_{\ell}=\begin{pmatrix}
e^{i\eta_{\ell_{1}}}& 0 & 0 \\
0 & e^{i\eta_{\ell_{2}}} & 0 \\
0 & 0 & e^{i\eta_{\ell_{3}}}
\end{pmatrix},\qquad \bar{\mathbf{m}}_{\ell}=\begin{pmatrix}
\vert a_{\ell}\vert & \vert b_{\ell}\vert & 0 \\
\vert b_{\ell}\vert & \vert a_{\ell}\vert & \vert c_{\ell}\vert \\
0 & \vert c_{\ell}\vert  & \vert g_{\ell}\vert 
\end{pmatrix}
\end{eqnarray}
with the following condition on the CP phases
\begin{equation}
\eta_{\ell_{1}}=\frac{arg(a_{\ell})}{2},\quad \eta_{\ell_{2}}=\frac{arg(a_{\ell})}{2},\quad \eta_{\ell_{3}}=\frac{arg(g_{\ell})}{2},\quad \eta_{\ell_{1}}+\eta_{\ell_{2}}=arg(b_{\ell}),\quad \eta_{\ell_{2}}+\eta_{\ell_{3}}=arg(c_{\ell}).
\end{equation} 

As a result of factorizing the CP violating phases, we have that $\mathbf{u}_{e L}=\mathbf{P}_{e}\mathbf{O}_{e}$, $\mathbf{u}_{e R}=\mathbf{P}^{\dagger}_{e}\mathbf{O}_{e}$ and  $\mathbf{u}_{\nu}=\mathbf{P}_{\nu}\mathbf{O}_{\nu}$. Let us obtain the orthogonal matrix that diagonalizes the real symmetric mass matrix, $\bar{\mathbf{m}}_{\ell}$. 

Here, we will consider two cases: the normal and inverted hierarchy in the neutrino masses.

\subsubsection{Normal Hierarchy (NH)}
For this case, the diagonalization procedure is valid for charged lepton and the active neutrinos. Then considering the mass matrix, $\bar{\mathbf{m}}_{\ell}$,  we can fix three free parameters in terms of the physical masses and unfixed one free parameter, $\vert a_{\ell}\vert$. This is,
\begin{eqnarray}
\vert g_{\ell}\vert&=& m_{\ell_{3}}-\vert m_{\ell_{2}}\vert+m_{\ell_{1}}-2\vert a_{\ell}\vert\nn\\
\vert b_{\ell}\vert&=&\sqrt{\frac{(m_{\ell_{3}}-\vert a_{\ell}\vert)(\vert m_{\ell_{2}}\vert+\vert a_{\ell}\vert)(m_{\ell{1}}-\vert a_{\ell}\vert)}{m_{\ell_{3}}-\vert m_{\ell_{2}}\vert+m_{\ell_{1}}-3\vert a_{\ell}\vert}}\nn\\
\vert c_{\ell}\vert&=&\sqrt{\frac{(m_{\ell_{3}}+m_{\ell_{1}}-2\vert a_{\ell}\vert)(m_{\ell_{3}}-\vert m_{\ell_{2}}\vert-2\vert a_{\ell}\vert)(\vert m_{\ell_{2}}\vert-m_{\ell_{1}}+2\vert a_{\ell}\vert)}{m_{\ell_{3}}-\vert m_{\ell_{2}}\vert +m_{\ell_{1}}-3\vert a_{\ell}\vert}}\label{fpno}
\end{eqnarray}
 where we have taken $m_{\ell_{2}}=-\vert m_{\ell_{2}}\vert$ in order to get real parameters. In addition, there is a constraint for the unfixed free parameter $m_{\ell_{3}}>\vert m_{\ell_{2}}\vert>m_{\ell_{1}}>\vert a_{\ell}\vert>0$. After a lengthy task, we obtain the orthogonal real matrix
\begin{equation}
\mathbf{O}_{\ell}=\begin{pmatrix}
\sqrt{\frac{( \tilde{m}_{\ell_{2}}+\tilde{a}_{\ell})(1-\tilde{a}_{\ell})\mathcal{M}_{2}}{\mathcal{D}_{1}}}& -\sqrt{\frac{(\tilde{m}_{\ell_{1}}-\tilde{a}_{\ell})(1-\tilde{a}_{\ell})\mathcal{M}_{1}}{\mathcal{D}_{2}}}
& \sqrt{\frac{(\tilde{m}_{\ell_{2}}+\tilde{a}_{\ell})(\tilde{m}_{\ell_{1}}-\tilde{a}_{\ell})\mathcal{M}_{3}}{\mathcal{D}_{3}}} 
\\ 
\sqrt{\frac{(\tilde{m}_{\ell_{1}}-\tilde{a}_{\ell})\mathcal{M}_{2}\mathcal{D}}{\mathcal{D}_{1}}}& \sqrt{\frac{(\tilde{m}_{\ell_{2}}+\tilde{a}_{\ell})\mathcal{M}_{1}\mathcal{D}}{\mathcal{D}_{2}}}
& \sqrt{\frac{(1-\tilde{a}_{\ell})\mathcal{M}_{3}\mathcal{D}}{\mathcal{D}_{3}}} \\ 
-\sqrt{\frac{(\tilde{m}_{\ell_{1}}-\tilde{a}_{\ell})\mathcal{M}_{1}\mathcal{M}_{3}}{\mathcal{D}_{1}}}&-\sqrt{\frac{(\tilde{m}_{\ell_{2}}+\tilde{a}_{\ell})\mathcal{M}_{2}\mathcal{M}_{3}}{\mathcal{D}_{2}}} 
& 
\sqrt{\frac{(1-\tilde{a}_{\ell})\mathcal{M}_{1}\mathcal{M}_{2}}{\mathcal{D}_{3}}}
\label{eq7}
\end{pmatrix} 
\end{equation}
with 
\begin{eqnarray}
\mathcal{M}_{1}&=&1+\tilde{m}_{\ell_{1}}-2\tilde{a}_{\ell},\quad 
\mathcal{M}_{2}=1-\tilde{m}_{\ell_{2}}-2\tilde{a}_{\ell},\quad
\mathcal{M}_{3}=\tilde{m}_{\ell_{2}}-\tilde{m}_{\ell_{1}}+2\tilde{a}_{\ell}
,\quad
\mathcal{D}=1-\tilde{m}_{\ell_{2}}+\tilde{m}_{\ell_{1}}-3\tilde{a}_{\ell};\nonumber\\
\mathcal{D}_{1}&=&(1-\tilde{m}_{\ell_{1}})( \tilde{m}_{\ell_{2}}+\tilde{m}_{\ell_{1}})\mathcal{D},\quad
\mathcal{D}_{2}=(1+\tilde{m}_{\ell_{2}})( \tilde{m}_{\ell_{2}}+\tilde{m}_{\ell_{1}})\mathcal{D},\quad
\mathcal{D}_{3}=(1+ \tilde{m}_{\ell_{2}})(1-\tilde{m}_{\ell_{1}})\mathcal{D},
\label{eq8}
\end{eqnarray}
where $\tilde{m}_{\ell_{2}}=\vert m_{\ell_{2}}\vert /m_{\ell_{3}}$, $\tilde{m}_{\ell_{1}}=m_{\ell_{1}}/m_{\ell_{3}}$ and $\tilde{a}_{\ell}=\vert a_{\ell}\vert/ m_{\ell_{3}}$. As we observed, for simplicity, the mixing matrix elements have been normalized by the heaviest mass. Therefore, the constraint is replaced by $1>\tilde{m}_{\ell_{2}}>\tilde{m}_{\ell_{1}}>\tilde{a}_{\ell}>0$. 

\subsubsection{Inverted Hierarchy (IH)}
For this ordering, we obtain the fixed free parameters
\begin{eqnarray}
\vert d_{\nu}\vert&=& m_{2}-\vert m_{1}\vert+m_{3}-2\vert a_{\nu}\vert\nn\\
\vert b_{\nu}\vert&=&\sqrt{\frac{(m_{3}-\vert a_{\nu}\vert)(\vert m_{1}\vert+\vert a_{\nu}\vert)(m_{2}-\vert a_{\nu}\vert)}{m_{2}-\vert m_{1}\vert+m_{3}-3\vert a_{\nu}\vert}}\nn\\
\vert c_{\nu}\vert&=&\sqrt{\frac{(\vert m_{1}\vert-m_{3}+2\vert a_{\nu}\vert)(m_{2}+ m_{3}-2\vert a_{\nu}\vert)(m_{2}-\vert m_{1}\vert-2\vert a_{\nu}\vert)}{m_{2}-\vert m_{1}\vert +m_{3}-3\vert a_{\nu}\vert}}\label{fpio}
\end{eqnarray}
where we have taken $m_{1}=-\vert m_{1}\vert$ for getting the real parameters. Therefore, the orthogonal real matrix is given by
\begin{equation}
\mathbf{O}_{\nu}=\begin{pmatrix}
-\sqrt{\frac{(1-\tilde{a}_{\nu})(\tilde{m}_{3}-\tilde{a}_{\nu})\mathcal{N}_{2}}{\mathcal{D}_{\nu_{1}}}}& \sqrt{\frac{(\tilde{m}_{{1}}+\tilde{a}_{\nu})(\tilde{m}_{3}-\tilde{a}_{\nu})\mathcal{N}_{1}}{\mathcal{D}_{\nu_{2}}}}
& \sqrt{\frac{(1-\tilde{a}_{\nu})(\tilde{m}_{1}+\tilde{a}_{\nu})\mathcal{N}_{3}}{\mathcal{D}_{\nu_{3}}}} 
\\ 
\sqrt{\frac{(\tilde{m}_{1}+\tilde{a}_{\nu})\mathcal{N}_{2}\mathcal{D}_{\nu}}{\mathcal{D}_{\nu_{1}}}}& \sqrt{\frac{(1-\tilde{a}_{\nu})\mathcal{N}_{1}\mathcal{D}_{\nu}}{\mathcal{D}_{\nu_{2}}}}
& \sqrt{\frac{(\tilde{m}_{3}-\tilde{a}_{\nu})\mathcal{N}_{3}\mathcal{D}_{\nu}}{\mathcal{D}_{\nu_{3}}}} \\ 
-\sqrt{\frac{(\tilde{m}_{1}+\tilde{a}_{\nu})\mathcal{N}_{1}\mathcal{N}_{3}}{\mathcal{D}_{\nu_{1}}}}&\sqrt{\frac{(1-\tilde{a}_{\nu})\mathcal{N}_{2}\mathcal{N}_{3}}{\mathcal{D}_{\nu_{2}}}} 
& 
-\sqrt{\frac{(\tilde{m}_{3}-\tilde{a}_{\nu})\mathcal{N}_{1}\mathcal{N}_{2}}{\mathcal{D}_{\nu_{3}}}}
\label{eq88}
\end{pmatrix} 
\end{equation}
where
\begin{eqnarray}
\mathcal{N}_{1}&=&\tilde{m}_{1}-\tilde{m}_{3}+2\tilde{a}_{\nu},\quad 
\mathcal{N}_{2}=1+\tilde{m}_{3}-2\tilde{a}_{\nu},\quad
\mathcal{N}_{3}=1-\tilde{m}_{1}-2\tilde{a}_{\nu}
,\quad
\mathcal{D}_{\nu}=1-\tilde{m}_{1}+\tilde{m}_{3}-3\tilde{a}_{\nu};\nonumber\\
\mathcal{D}_{\nu_{1}}&=&(1+\tilde{m}_{1})( \tilde{m}_{1}+\tilde{m}_{3})\mathcal{D}_{\nu},\quad
\mathcal{D}_{\nu_{2}}=(1+\tilde{m}_{1})( 1-\tilde{m}_{3})\mathcal{D}_{\nu},\quad
\mathcal{D}_{\nu_{3}}=(1- \tilde{m}_{3})(\tilde{m}_{1}+\tilde{m}_{3})\mathcal{D}_{\nu},
\label{eq99}
\end{eqnarray}
where $\tilde{m}_{{1}}=\vert m_{{1}}\vert /m_{2}$, $\tilde{m}_{3}=m_{3}/m_{2}$ and $\tilde{a}_{\nu}=\vert a_{\nu}\vert/ m_{2}$. In this parametrization, there is a constraint among the neutrino masses and the free parameter $\tilde{a}_{\nu}$, this is $1>\tilde{m}_{1}>\tilde{m}_{3}>\tilde{a}_{\nu}>0$.

Hence, we end up having the PMNS mixing matrix $\mathbf{V}^{i}=\mathbf{U}^{\dagger}_{e L}\mathbf{U}^{i}_{\nu}=\mathbf{O}^{T}_{e}\bar{\mathbf{P}}_{e}\mathbf{O}^{i}_{\nu}$ with $i= NH, IH$. In addition, $\bar{\mathbf{P}}_{e}=\mathbf{P}^{\dagger}_{e}\mathbf{P}_{\nu}\equiv\textrm{Diag.}(1, 1,e^{i\eta_{\nu}})$ with $\eta_{\nu}=\eta_{\nu_{3}}-\eta_{\tau}$. Thus, we can compare our expression with the standard parametrization of the PMNS mixing matrix such that the reactor, atmospheric and solar angles are well determined by
\begin{eqnarray}\label{mixan}
\sin{\theta}_{13}&=&\vert (\mathbf{V}^{i})_{13}\vert=\vert (\mathbf{O}_{e})_{11} (\mathbf{O}^{i}_{\nu})_{13}+(\mathbf{O}_{e})_{21} (\mathbf{O}^{i}_{\nu})_{23}+(\mathbf{O}_{e})_{31} (\mathbf{O}^{i}_{\nu})_{33}e^{i\eta_{\nu}}
\vert, \nn\\
\sin{\theta}_{23}&=&\frac{\vert (\mathbf{V}^{i})_{23}\vert}{\sqrt{1-\sin^{2}{\theta}_{13}}}=\frac{\vert  (\mathbf{O}_{e})_{12} (\mathbf{O}^{i}_{\nu})_{13}+(\mathbf{O}_{e})_{22} (\mathbf{O}^{i}_{\nu})_{23}+(\mathbf{O}_{e})_{32} (\mathbf{O}^{i}_{\nu})_{33}e^{i\eta_{\nu}} \vert}{\sqrt{1-\sin^{2}{\theta}_{13}}},\nn\\
\sin{\theta}_{12}&=&\frac{\vert (\mathbf{V}^{i})_{12}\vert}{\sqrt{1-\sin^{2}{\theta}_{13}}}=\frac{\vert (\mathbf{O}_{e})_{11} (\mathbf{O}^{i}_{\nu})_{12}+(\mathbf{O}_{e})_{21} (\mathbf{O}^{i}_{\nu})_{22}+(\mathbf{O}_{e})_{31} (\mathbf{O}^{i}_{\nu})_{32}e^{i\eta_{\nu}}
\vert}{\sqrt{1-\sin^{2}{\theta}_{13}}}.
\end{eqnarray}

In the PMNS matrix there are three free parameters namely: $\vert a_{e}\vert$, $\vert a_{\nu}\vert$ and one CP violating phase $\eta_{\nu}$, in Eq.(\ref{mixan}). In fact, due to of lacking information on the absolute neutrino masses, the lightest one may be considered as an extra free parameter.

On the other hand, we would like to point out a little comment on the Majorana phases for each hierarchy. We have considered the CP parities for the complex neutrino masses which means that these can be either $0$ or $\pi$. Thus, for the normal and inverted ordering we have $(m_{3}, m_{2}, m_{1})=(+,-,+)$ and $(m_{3}, m_{2}, m_{1})=(+,+,-)$, respectively. Those CP parities values ensure that the fixed parameters given in Eq. (\ref{fpno}) and Eq.(\ref{fpio}) are reals.

\section{Results}

\subsection{Analytical study}

In order to try of figuring out the allowed region for free parameters, let us make a brief analytical study on the mixing angles formulas. To do so, we have to keep in mind that for the normal and inverted hierarchy, two neutrino masses can be fixed in terms of the squared mass scales and the lightest neutrino mass. This is,
\begin{eqnarray}\label{massfix}
m_{3}&=&\sqrt{\Delta m^{2}_{31}+m^{2}_{1}},\qquad \vert m_{2}\vert=\sqrt{\Delta m^{2
	}_{21}+m^{2}_{1}},\qquad \textrm{Normal Hierarchy}\nn\\
m_{2}&=&\sqrt{\Delta m^{2}_{13}+\Delta m^{2}_{21}+m^{2}_{3}},\qquad \vert m_{1}\vert =\sqrt{\Delta m^{2
	}_{13}+m^{2}_{3}}.\qquad \textrm{Inverted Hierarchy}
\end{eqnarray}

In addition, the experimental data, that will be used in this analytical and numerical study, is given in the table \ref{Tabpdga}

\begin{table}[th]
	\begin{center}
		\begin{tabular}{|c|c|c|}
			\hline\hline
			Observable & Experimental value \\ \hline
			$m_{e}(MeV)$ & \quad $0.5109989461 \pm 0.0000000031$ \\ \hline
			$m_{\mu}(MeV)$ &  \quad $ 105.6583745 \pm 0.0000024$ \\ \hline
			$m_{\tau}(MeV)$ &  \quad $1776.86 \pm 0.12$ \\ \hline
			$\frac{\Delta m^{2}_{21}}{10^{-5}~eV^{2}}$ & \quad $7.50_{-0.20}^{+0.22}$ \\ \hline
			$\frac{\Delta m^{2}_{31}}{10^{-3}~eV^{2}}$ & \quad $2.55_{-0.03}^{+0.02}$ \quad ($2.45_{-0.03}^{+0.02}$) \\ \hline
			$\sin^{2}{\theta}_{12}$ & \quad $0.318 \pm 0.16 $ \\ \hline
			$\sin^{2}{\theta}_{23}$ &  \quad $0.574 \pm 0.14  $ ($0.578^{+0.10}_{-0.17}$)\\ \hline
			$\sin^{2}{\theta}_{13}$ & \quad $0.02200^{+0.069}_{-0.062}$ ($0.02225^{+0.064}_{-0.070}$)\\ \hline
			$\delta_{CP}/^{\circ}$ & \quad $ 194^{+24}_{-22}$ ($ 284^{+26}_{-28}$) \\ \hline\hline
		\end{tabular}%
	\end{center}
	\caption{Experimental values of the lepton masses and PMNS parameters~\cite{deSalas:2020pgw, Zyla:2020zbs}.}
	\label{Tabpdga}
\end{table}

In the current analysis, central values will be used for the normalized masses and there is a hierarchy among those, this is,  $\tilde{m}_{\mu}>\tilde{m}_{e}/ \tilde{m}_{\mu}>\tilde{m}_{e}$, $\tilde{m}_{2}>\tilde{m}_{1}/ \tilde{m}_{2}>\tilde{m}_{1}$ (for normal ordering) and $\tilde{m}_{1}>\tilde{m}_{3}/ \tilde{m}_{1}\gtrsim\tilde{m}_{3}$ (for inverted ordering); actually, for the last hierarchy we have $m_{2}\approx m_{1}(1+ \Delta m^{2}_{21}/2m^{2}_{1})$, then $\tilde{m}_{3}\approx \tilde{m}_{3}/ \tilde{m}_{1}$. Consequently, we get the following values
$\tilde{m}_{e}\approx 2.9\times 10^{-4}$, $\tilde{m}_{e}/ \tilde{m}_{\mu}\approx 4.8\times 10^{-3}$ and $\tilde{m}_{\mu} \approx 5.9\times 10^{-2}$. At the same time, for the neutrinos one obtains
\begin{itemize}
	\item Normal Hierarchy
	\begin{equation}\label{normasses}
\tilde{m}_{1}\approx 2\times 10^{-2};\qquad \frac{\tilde{m}_{1}}{\tilde{m}_{2}}\approx 0.115,\qquad \tilde{m}_{2}\approx 0.173. 
\end{equation}
with $m_{1}\approx 0.001$ for the lightest mass.

\item Inverted Hierarchy
\begin{equation}
\tilde{m}_{3}\approx 0.195;\qquad \frac{\tilde{m}_{3}}{\tilde{m}_{1}}\approx 0.198,\qquad \tilde{m}_{1}\approx 1 
\end{equation}
with $m_{3}\approx 0.01$. 
\end{itemize}
 
Notice that particular values for the lightest neutrino mass have been considered for the normal and inverted hierarchy. Thus, we will obtain approximately the matrices $\mathbf{O}_{e}$ and $\mathbf{O}_{\nu}$ for the normal and inverted ordering, then the mixing angles must be calculated in analytical way for different scenarios.

\begin{description}
\item [Normal Hierarchy]
 ($1>\tilde{m}_{\ell_{2}}>\tilde{m}_{\ell_{1}}>\tilde{a}_{\ell}>0$).

\begin{itemize}
\item Case I: $\tilde{a}_{\ell}\approx 0$. In this limit, the Fritzsch textures are recovered and the orthogonal matrix is given by
	
\begin{equation}
\mathbf{O}_{\ell}\approx\begin{pmatrix}
\sqrt{\frac{\tilde{m}_{\ell_{2}}(1-\tilde{m}_{\ell_{2}})}{(1-\tilde{m}_{\ell_{1}})(\tilde{m}_{\ell_{2}}+\tilde{m}_{\ell_{1}})(1-\tilde{m}_{\ell_{2}}+\tilde{m}_{\ell_{1}})}}& -\sqrt{\frac{\tilde{m}_{\ell_{1}}(1+\tilde{m}_{\ell_{1}})}{(1+\tilde{m}_{\ell_{2}})(\tilde{m}_{\ell_{2}}+\tilde{m}_{\ell_{1}})(1-\tilde{m}_{\ell_{2}}+\tilde{m}_{\ell_{1}})}}
& \sqrt{\frac{\tilde{m}_{\ell_{2}}\tilde{m}_{\ell_{1}}(\tilde{m}_{\ell_{2}}-\tilde{m}_{\ell_{1}})}{(1-\tilde{m}_{\ell_{1}})(1+\tilde{m}_{\ell_{2}})(1-\tilde{m}_{\ell_{2}}+\tilde{m}_{\ell_{1}})}}
	\\ 
	\sqrt{\frac{\tilde{m}_{\ell_{1}}(1-\tilde{m}_{\ell_{2}})}{(1-\tilde{m}_{\ell_{1}})(\tilde{m}_{\ell_{2}}+\tilde{m}_{\ell_{1}})}}& \sqrt{\frac{\tilde{m}_{\ell_{2}}(1+\tilde{m}_{\ell_{1}})}{(\tilde{m}_{\ell_{2}}+\tilde{m}_{\ell_{1}})(1+\tilde{m}_{\ell_{2}})}}
	& \sqrt{\frac{(\tilde{m}_{\ell_{2}}-\tilde{m}_{\ell_{1}})}{(1-\tilde{m}_{\ell_{1}})(1+\tilde{m}_{\ell_{2}})}}
	\\ 
	-\sqrt{\frac{\tilde{m}_{\ell_{1}}(\tilde{m}_{\ell_{2}}-\tilde{m}_{\ell_{1}})(1+\tilde{m}_{\ell_{1}})}{(1-\tilde{m}_{\ell_{1}})(\tilde{m}_{\ell_{2}}+\tilde{m}_{\ell_{1}})(1-\tilde{m}_{\ell_{2}}+\tilde{m}_{\ell_{1}})}}&-\sqrt{\frac{\tilde{m}_{\ell_{2}}(\tilde{m}_{\ell_{2}}-\tilde{m}_{\ell_{1}})(1-\tilde{m}_{\ell_{2}})}{(1+\tilde{m}_{\ell_{2}})(\tilde{m}_{\ell_{2}}+\tilde{m}_{\ell_{1}})(1-\tilde{m}_{\ell_{2}}+\tilde{m}_{\ell_{1}})}}
	& \sqrt{\frac{(1-\tilde{m}_{\ell_{2}})(1+\tilde{m}_{\ell_{1}})}{(1-\tilde{m}_{\ell_{1}})(1+\tilde{m}_{\ell_{2}})(1-\tilde{m}_{\ell_{2}}+\tilde{m}_{\ell_{1}})}}
	\end{pmatrix}. \label{sce1}
	\end{equation}

\item Case II: $\tilde{a}_{\ell}\approx\tilde{m}_{\ell_{1}}$. 
\begin{equation}
\mathbf{O}_{\ell}\approx \begin{pmatrix}
1 & 0 & 0 \\
0 & \sqrt{\frac{1-\tilde{m}_{\ell_{1}}}{1+\tilde{m}_{\ell_{2}}}} & \sqrt{\frac{\tilde{m}_{\ell_{2}}+\tilde{m}_{\ell_{1}}}{1+\tilde{m}_{\ell_{2}}}} \\
0 & -\sqrt{\frac{\tilde{m}_{\ell_{2}}+\tilde{m}_{\ell_{1}}}{1+\tilde{m}_{\ell_{2}}}} & \sqrt{\frac{1-\tilde{m}_{\ell_{1}}}{1+\tilde{m}_{\ell_{2}}}}
\end{pmatrix}.\label{sce2}
\end{equation}	
\end{itemize}

\item [Inverted Hierarchy] ($1>\tilde{m}_{1}>\tilde{m}_{3}>\tilde{a}_{\nu}>0$). 

\begin{itemize}	
\item Case I: $\tilde{a}_{\nu}\approx 0$. 

\begin{equation}
\mathbf{O}_{\nu}\approx\begin{pmatrix}
-\sqrt{\frac{\tilde{m}_{3}(1+\tilde{m}_{3})}{(1+\tilde{m}_{1})(\tilde{m}_{1}+\tilde{m}_{3})(1-\tilde{m}_{1}+\tilde{m}_{3})}}& \sqrt{\frac{\tilde{m}_{1}\tilde{m}_{3}(1- \tilde{m}_{3})}{(1+\tilde{m}_{1})(1-\tilde{m}_{3})(1-\tilde{m}_{1}+\tilde{m}_{3})}}
& -\sqrt{\frac{\tilde{m}_{1}(1-\tilde{m}_{1})}{(1-\tilde{m}_{3})(\tilde{m}_{1}+\tilde{m}_{3})(1-\tilde{m}_{1}+\tilde{m}_{3})}}
\\ 
\sqrt{\frac{\tilde{m}_{1}(1+\tilde{m}_{3})}{(1+\tilde{m}_{1})(\tilde{m}_{1}+\tilde{m}_{3})}}& \sqrt{\frac{\tilde{m}_{1}-\tilde{m}_{3}}{(1+\tilde{m}_{1})(1-\tilde{m}_{3})}}
& \sqrt{\frac{\tilde{m}_{3}(1-\tilde{m}_{1})}{(1-\tilde{m}_{3})(\tilde{m}_{1}+\tilde{m}_{3})}}
\\ 
-\sqrt{\frac{\tilde{m}_{1}(\tilde{m}_{1}-\tilde{m}_{3})(1-\tilde{m}_{1})}{(1+\tilde{m}_{1})(\tilde{m}_{1}+\tilde{m}_{3})(1-\tilde{m}_{1}+\tilde{m}_{3})}}&\sqrt{\frac{(1+\tilde{m}_{3})(1-\tilde{m}_{1})}{(1+\tilde{m}_{1})(1-\tilde{m}_{3})(1-\tilde{m}_{1}+\tilde{m}_{3})}}
& -\sqrt{\frac{\tilde{m}_{3}(\tilde{m}_{1}-\tilde{m}_{3})(1+\tilde{m}_{3})}{(1-\tilde{m}_{3})(\tilde{m}_{1}+\tilde{m}_{3})(1-\tilde{m}_{1}+\tilde{m}_{3})}}
\end{pmatrix}. \label{sce3}
\end{equation}

\item Case II: $\tilde{a}_{\nu}\approx\tilde{m}_{3}$. 

\begin{equation}
\mathbf{O}_{\nu}\approx\begin{pmatrix}
0 & 0
& 1
\\ 
\sqrt{\frac{1-\tilde{m}_{3}}{1+\tilde{m}_{1}}}& \sqrt{\frac{\tilde{m}_{1}+\tilde{m}_{3}}{1+\tilde{m}_{1}}}
& 0
\\ 
-\sqrt{\frac{\tilde{m}_{1}+\tilde{m}_{3}}{1+\tilde{m}_{1}}} & \sqrt{\frac{1-\tilde{m}_{3}}{1+\tilde{m}_{1}}}
& 0
\end{pmatrix}.\label{sce4}
\end{equation}
\end{itemize}
\end{description}

Having obtained the above approximated matrices, then we can obtain the mixing angles for different scenarios and some combinations:
\begin{enumerate}
	\item Normal hierarchy

\begin{itemize}
	\item Scenario A: If $\mathbf{O}_{e}$ and $\mathbf{O}_{\nu}$ were like Eqn. (\ref{sce1}), then the mixing angles would be 
	\begin{eqnarray}
	\sin{\theta_{13}}&\approx& \lvert \tilde{m}_{2}\sqrt{\tilde{m}_{1}\left(1-\frac{\tilde{m}_{1}}{\tilde{m}_{2}}\right)}+\sqrt{\frac{\tilde{m}_{e}}{\tilde{m}_{\mu}}}\sqrt{\tilde{m}_{2}\left(1-\tilde{m}_{2}\right)}-\sqrt{\tilde{m}_{e}}\sqrt{1-\tilde{m}_{2}}~ e^{i\eta_{\nu}} \rvert;\notag\\
	\sin{\theta_{23}}&\approx&\lvert \frac{-\sqrt{\frac{\tilde{m}_{e}}{\tilde{m}_{\mu}}}~\tilde{m}_{2}\sqrt{\tilde{m}_{1}\left(1-\frac{\tilde{m}_{1}}{\tilde{m}_{2}}\right)}+\sqrt{\tilde{m}_{2}\left(1-\tilde{m}_{2}\right)}-\sqrt{\tilde{m}_{\mu}}\sqrt{1-\tilde{m}_{2}}~ e^{i\eta_{\nu}}}{\sqrt{1-\sin^{2}{\theta_{13}}}}\rvert;\notag\\
	\sin{\theta_{12}}&\approx&\lvert \frac{-\sqrt{\frac{\tilde{m}_{1}}{\tilde{m}_{2}}\left(1-\frac{\tilde{m}_{1}}{\tilde{m}_{2}}\right)}+\sqrt{\frac{\tilde{m}_{e}}{\tilde{m}_{\mu}}}\sqrt{1-\tilde{m}_{2}}+\sqrt{\tilde{m}_{e}}~\sqrt{\tilde{m}_{2}\left(1-\tilde{m}_{2}\right)}~ e^{i\eta_{\nu}}}{\sqrt{1-\sin^{2}{\theta_{13}}}}\rvert.
	\end{eqnarray}
where the notable hierarchy in the charged lepton has been taken into account. In the above expressions, the reactor, atmospheric and solar angles are controlled by the ratio $\sqrt{\tilde{m}_{e}/\tilde{m}_{\mu}}\approx0.069$, $\sqrt{\tilde{m}_{2}}\approx 0.41$ and $\sqrt{\tilde{m}_{1}/\tilde{m}_{2}}\approx0.34$, respectively.  In order to enhance the angle values, the phase $\eta_{\nu}$ must be near to $\pi$. In this way, we have that $\sin{\theta_{13}}\approx0.06$, $\sin{\theta_{23}}\approx 0.6$ and $\sin{\theta_{12}}\approx 0.25$. As result of this, the reactor and solar angle are not in the allowed experimental region with the neutrino masses values given in Eq. (\ref{normasses}).
	\item Scenario B. If $\mathbf{O}_{e}$ and $\mathbf{O}_{\nu}$ were like Eqn. (\ref{sce2}), then 
	one would get
    \begin{eqnarray}
	\sin{\theta_{13}}&\approx& 0;\notag\\
	\sin{\theta_{23}}&\approx&\lvert \sqrt{\frac{1-\tilde{m}_{e}}{1+\tilde{m}_{\mu}}}~ \sqrt{\frac{\tilde{m}_{2}+\tilde{m}_{1}}{1+\tilde{m}_{2}}}-\sqrt{\frac{\tilde{m}_{\mu}+\tilde{m}_{e}}{1+\tilde{m}_{\mu}}}
	~\sqrt{\frac{1-\tilde{m}_{1}}{1+\tilde{m}_{2}}}~ e^{i\eta_{\nu}}\rvert;\notag\\
	\sin{\theta_{12}}&\approx&0.
	\end{eqnarray}
	So that this case is completely discarded.
	
	\item Scenario C: If $\mathbf{O}_{\nu}$ and $\mathbf{O}_{e}$ were like Eqs. (\ref{sce1}) and (\ref{sce2}) respectively, then the mixing angles would be
	\begin{eqnarray}
	\sin{\theta_{13}}&\approx& \lvert \tilde{m}_{2} \sqrt{\tilde{m}_{1}(1-\frac{\tilde{m}_{1}}{\tilde{m}_{2}})} \rvert;\notag\\
	\sin{\theta_{23}}&\approx&\lvert \frac{ \sqrt{\frac{1-\tilde{m}_{e}}{1+\tilde{m}_{\mu}}}
		~\sqrt{\tilde{m}_{2}(1-\tilde{m}_{2})}- \sqrt{\frac{\tilde{m}_{\mu}+\tilde{m}_{e}}{1+\tilde{m}_{\mu}}}
		~\sqrt{1-\tilde{m}_{2}}~ e^{i\eta_{\nu}}}{\sqrt{1-\sin^{2}{\theta_{13}}}}\rvert;\notag\\
	\sin{\theta_{12}}&\approx& \lvert \frac{ \sqrt{\frac{\tilde{m}_{1}}{\tilde{m}_{2}}(1-\frac{\tilde{m}_{1}}{\tilde{m}_{2}})}}{\sqrt{1-\sin^{2}{\theta_{13}}}}\rvert
	\end{eqnarray}
	
As one can notice, the reactor angle is tiny in comparison to the scenario A, the atmospheric and solar angle are handled by the 
$\sqrt{\tilde{m}_{2}}\approx 0.41$ and $\sqrt{\tilde{m}_{1}/\tilde{m}_{2}}\approx0.34$; the atmospheric angle value can be increased by allowing that the phase $\eta_{\nu}$ must be $\pi$. Therefore, we obtain $\sin{\theta_{13}}\approx0.016$, $\sin{\theta_{23}}\approx 0.58$ and $\sin{\theta_{12}}\approx 0.32$.

\item Scenario D: If $\mathbf{O}_{\nu}$ and $\mathbf{O}_{e}$ were like Eq.(\ref{sce2}) and Eq. (\ref{sce1}) respectively, then one would obtain
	
	\begin{eqnarray}
	\sin{\theta_{13}}&\approx& \lvert \sqrt{\frac{\tilde{m}_{e}}{\tilde{m}_{\mu}}}~\sqrt{\frac{\tilde{m}_{2}+\tilde{m}_{1}}{1+\tilde{m}_{2}}}-\sqrt{\tilde{m}_{e}}~\sqrt{\frac{1-\tilde{m}_{1}}{1+\tilde{m}_{2}}}~e^{i\eta_{\nu}} \rvert;\notag\\
	\sin{\theta_{23}}&\approx&\lvert \frac{ \sqrt{\frac{\tilde{m}_{2}+\tilde{m}_{1}}{1+\tilde{m}_{2}}}-\sqrt{\tilde{m}_{\mu}}~
		~\sqrt{\frac{1-\tilde{m}_{1}}{1+\tilde{m}_{2}}}~ e^{i\eta_{\nu}}}{\sqrt{1-\sin^{2}{\theta_{13}}}}\rvert;\notag\\
	\sin{\theta_{12}}&\approx& \lvert \frac{ \sqrt{\frac{\tilde{m}_{e}}{\tilde{m}_{\mu}}}~ \sqrt{\frac{1-\tilde{m}_{1}}{1+\tilde{m}_{2}}}-\sqrt{\tilde{m}_{e}}
		~\sqrt{\frac{\tilde{m}_{2}+\tilde{m}_{1}}{1+\tilde{m}_{2}}}~ e^{i\eta_{\nu}}}{\sqrt{1-\sin^{2}{\theta_{13}}}}\rvert
	\end{eqnarray}
In this scenario, the reactor angle is smaller (larger) than scenario A (C); the solar angle is smaller than the scenarios A and C so that this case is rule out.	
\end{itemize}

\item Inverted hierarchy

\begin{itemize}
	\item Scenario E: If the charged lepton and the neutrino mixing matrices were like Eq.(\ref{sce1}) and  Eq.(\ref{sce3}), then the observables would be
	\begin{eqnarray}
	\sin{\theta_{13}}&\approx&  \sqrt{\tilde{m}_{e}};\notag\\
	\sin{\theta_{23}}&\approx& \sqrt{\tilde{m}_{\mu}};\notag\\
	\sin{\theta_{12}}&\approx&  \frac{1}{\sqrt{2}}\left(1+\sqrt{\frac{\tilde{m}_{e}}{\tilde{m}_{\mu}}}\right).
	\end{eqnarray}
This scenario is discarded since that the reactor and atmospheric angles are tiny.	
	
\item Scenario F: If the charged lepton  and the neutrino mixing matrices were like Eq.(\ref{sce2}) and  Eq.(\ref{sce3}), then  the observable would be
	
	\begin{eqnarray}
	\sin{\theta_{13}}&\approx& 0;\notag\\
	\sin{\theta_{23}}&\approx& \sqrt{\frac{\tilde{m}_{\mu}+\tilde{m}_{e}}{1+\tilde{m}_{\mu}}};\notag\\
	\sin{\theta_{12}}&\approx&  \frac{1}{\sqrt{2}}.
	\end{eqnarray}
Analogously to the previous case, this scenario is ruled out by the predictions on the reactor and atmospheric angles which come out being small.	

\item Scenario G: $\mathbf{O}_{\nu}$ and $\mathbf{O}_{e}$ given by  Eqs. (\ref{sce4}) and (\ref{sce1}), respectively, then
\begin{eqnarray}
\sin{\theta_{13}}&\approx& 1;\notag\\
\sin{\theta_{23}}&>&1 ;\notag\\
\sin{\theta_{12}}&>&1.
\end{eqnarray}	

\item Scenario H: $\mathbf{O}_{\nu}$ and $\mathbf{O}_{e}$ given by  Eqs. (\ref{sce4}) and (\ref{sce2}), respectively, then

\begin{eqnarray}
\sin{\theta_{13}}&\approx& 1;\notag\\
\sin{\theta_{23}}&>& 1;\notag\\
\sin{\theta_{12}}&>& 1.
\end{eqnarray}	
The last two scenarios are completely ruled out due to the reactor angle is close to $1$.
\end{itemize}
\end{enumerate}

As consequence of this analytical study, speaking roughly there are two scenarios (A and C) which seem to provide allowed values for the observables. Let us add that one would expect changes in the mentioned scenarios when
the lightest neutrino mass varies in its allowed region.

\subsection{Numerical study}

The numerical analysis  consists of scattered plots to constrain the allowed region for each free parameters. Then, we will be working with the following expressions 
\begin{eqnarray}\label{mixan2}
\sin^{2}{\theta}_{13}&=&\sin^{2}{\theta}_{13}\left(\vert a_{e}\vert, \vert a_{\nu}\vert, \eta_{\nu}, m_{j} \right)\nn\\
\sin^{2}{\theta}_{23}&=&\sin^{2}{\theta}_{23}\left(\vert a_{e}\vert, \vert a_{\nu}\vert, \eta_{\nu}, m_{j} \right) \nn\\
\sin^{2}{\theta}_{12}&=&\sin^{2}{\theta}_{12}\left(\vert a_{e}\vert, \vert a_{\nu}\vert, \eta_{\nu}, m_{j} \right)
\end{eqnarray}
where $m_{j}$ with $j=1, 3$ stands for the lightest neutrino mass for normal and inverted hierarchy, respectively. 

In the scattered plots, we will vary the free parameters in such a way those satisfy their respective constraints. For the lightest neutrino mass, in the normal (inverted) case, we have $1>\tilde{m}_{2}>\tilde{m}_{1}>\tilde{a}_{\nu}>0$  ($1>\tilde{m}_{1}>\tilde{m}_{3}>\tilde{a}_{\nu}>0$); along with this, for each hierarchy, the lightest mass varies in the region $0-0.9~eV$, the effective phase $2\pi\geq \eta_{\nu}\geq0$ and the charged lepton parameter $1>\tilde{m}_{\mu}>\tilde{m}_{e}>\tilde{a}_{e}>0$. Then, we demand that our theoretical expressions satisfy the experimental bounds up to $3\sigma$, this allows us to scan the allowed regions for the free parameters that fit quite well the experimental results. Finally, as a model prediction, the $\delta_{CP}$ CP-violating phase and the effective Majorana neutrino mass are fitted.

\begin{figure}[h]
	\centering
	\includegraphics[scale=0.4]{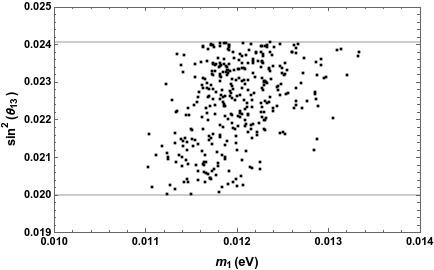}
	\hspace{1mm}\includegraphics[scale=0.4]{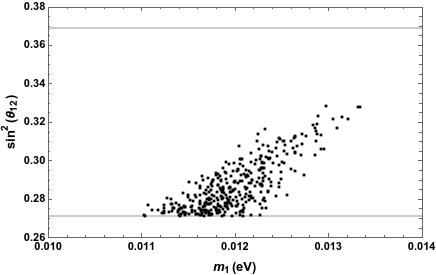}\\
	\includegraphics[scale=0.4]{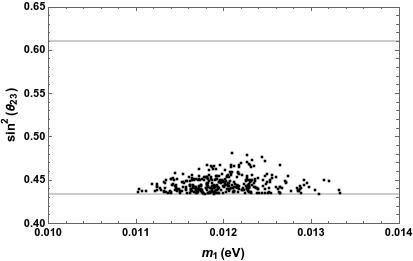}\hspace{1mm}\includegraphics[scale=0.4]{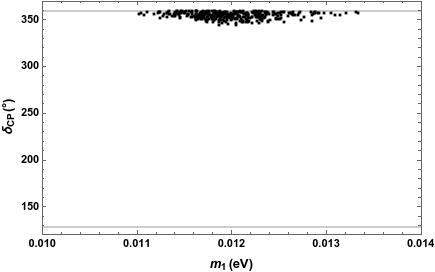}
	\caption{From left to right: the reactor, solar, atmospheric angles and CP phase versus the lightest neutrino mass. The thick line stands for $3~\sigma$ of C. L.}\label{f1}
\end{figure}
In the Fig. (\ref{f1}), we observe that there is a region ($0.01-0.014$~eV) for the lightest neutrino mass where the observables are in great according to the experimental results.

\begin{figure}[h!]\centering
	\includegraphics[scale=0.4]{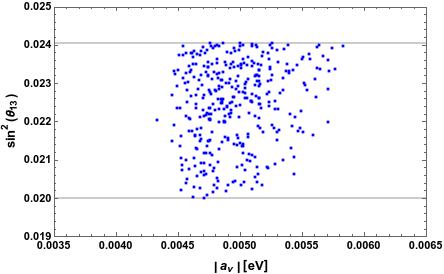}
	\hspace{1mm}\includegraphics[scale=0.4]{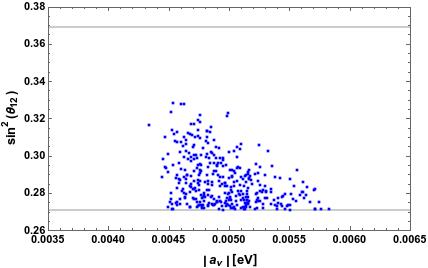}\\
	\includegraphics[scale=0.4]{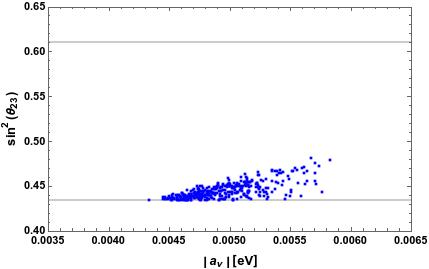}\hspace{1mm}\includegraphics[scale=0.4]{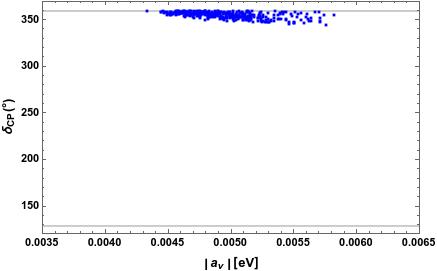}
	\caption{From left to right: the reactor, solar, atmospheric angles and CP phase versus the $\vert a_{\nu}\vert$ parameter. The thick line stands for $3~\sigma$ of C. L.}\label{f2}
\end{figure}

According to the Fig. (\ref{f2}), the $a_{\nu}~( \tilde{a}_{\nu})$ prefers small values for fitting the mixing angles. This means the Fritzsch textures are favored but a small deviation is necessary to accommodate the observables up to $3~\sigma$.

%\begin{figure}[h!]\centering
%	\includegraphics[scale=0.45]{BRVT.jpeg}
%	\hspace{10mm}\includegraphics[scale=0.45]{BRMT.jpeg}
%	\caption{From left to right: BR($\mu\rightarrow e\gamma$) versus the $v_{\Delta}$ and $m_{\Delta}$ parameter. The thick line stands for $3~\sigma$ of C. L.}\label{f6}
%\end{figure}

\begin{figure}[h!]\centering
	\includegraphics[scale=0.45]{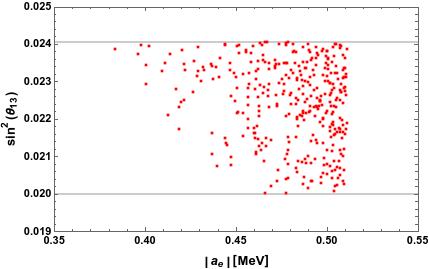}
	\hspace{1mm}\includegraphics[scale=0.45]{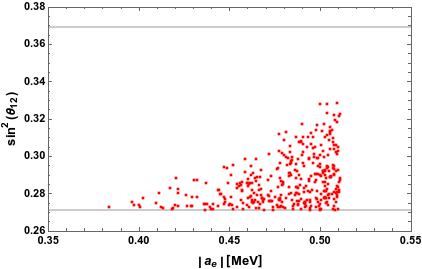}\\
	\includegraphics[scale=0.45]{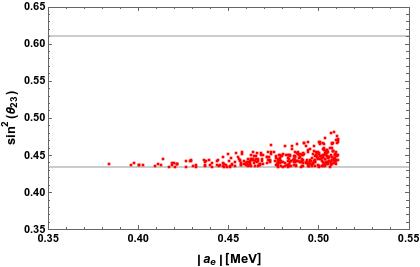}\hspace{1mm}\includegraphics[scale=0.45]{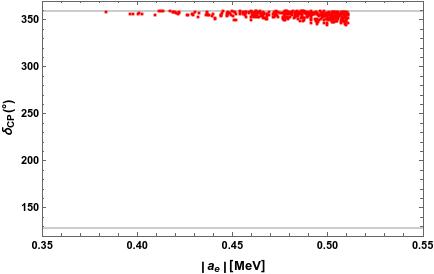}
	\caption{From left to right: the reactor, solar, atmospheric angles and CP phase versus the $\vert b_{e}\vert$ parameter. The thick line stands for $3~\sigma$ of C. L.}\label{f3}
\end{figure}

\begin{figure}[h!]\centering
	\includegraphics[scale=0.45]{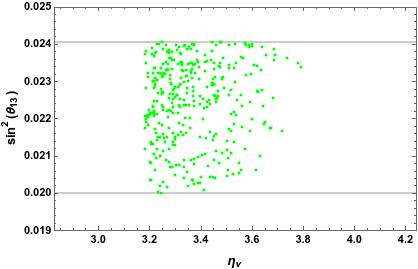}
	\hspace{1mm}\includegraphics[scale=0.45]{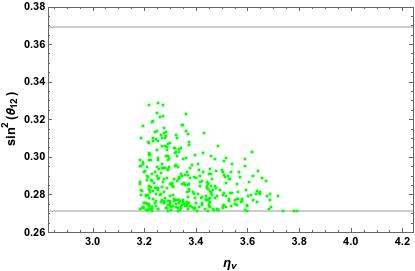}\\
	\includegraphics[scale=0.45]{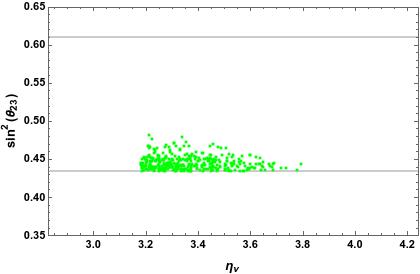}\hspace{1mm}\includegraphics[scale=0.45]{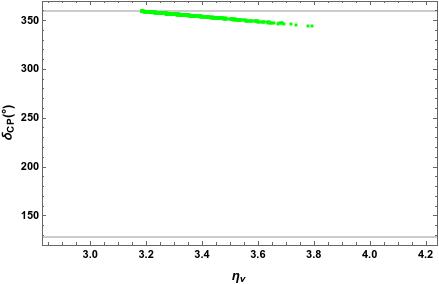}
	\caption{From left to right: the reactor, solar, atmospheric angles and CP phase versus the effective phase, $\eta_{\nu}$, parameter. The thick line stands for $3~\sigma$ of C. L.}\label{f4}
\end{figure}

In the charged lepton sector, the $a_{e}~(\tilde{a}_{e})$
parameter region is close to the electron mass as can be seen in Fig. (\ref{f3}), this is, $a_{e}\approx m_{e}$, so that the observables are well accommodated in the scenario C. Let us focus in the $\eta_{\nu}$ phase which lies in a region around $\pi$ value, the full region is shown in the Fig. (\ref{f4}).

To summarize, a set of free parameters has been found in which the reactor, solar and the atmospheric angles can accommodate quite well but this latter lies in the allowed low region ($3~\sigma$). In addition, the model predicts large values for the Dirac CP-violating phase which is close to the up region according to the experimental data.

\section{Model Predictions}

\subsection{Effective Majorana neutrino mass rate}

Going back to the comment about CP parities for the complex neutrino masses, we want to perform the effective Majorana mass of the electron neutrino, which is defined by

\begin{equation}
	\vert m_{ee}\vert=\vert m_{1} V^{2}_{e1}+ m_{2} V^{2}_{e2}+ m_{3} V^{2}_{e3}\vert,
\end{equation}

where $m_{i}$ and $V_{ei}$ ($i=1,2,3$) are the complex neutrino masses and PMNS matrix elements. As it is well known,  the lowest upper bound on $|m_{ee}|<0.22~eV$ was provided
by GERDA phase-I data \cite{Agostini:2013mzu} and this value has been significantly reduced by GERDA phase-II data \cite{Agostini:2017iyd}. 

\begin{figure}[h!]\centering
	\includegraphics[scale=0.45]{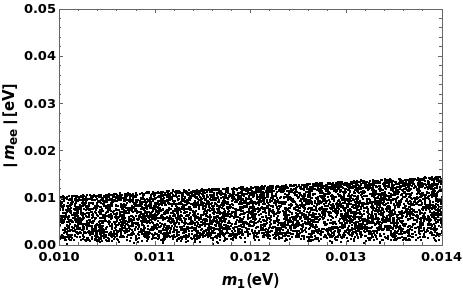}
	\hspace{1mm}\includegraphics[scale=0.45]{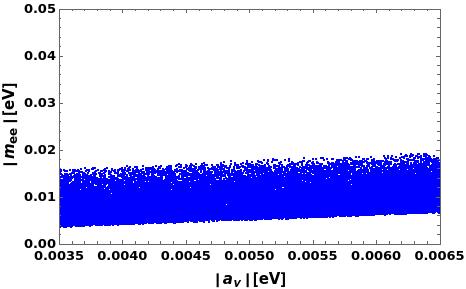}
	\caption{From left to right: $\vert m_{ee}\vert$ versus $m_{1}$ and $\vert a_{\nu} \vert$ parameters, respectively. These scattered plots correspond to the normal ordering where the CP parities for the complex neutrino masses are $(m_{3}, m_{2}, m_{1})=(+,-,+)$.}\label{f5}
\end{figure}

In the previous section, we found a set of values for the free parameters (see Fig. (\ref{f1}-\ref{f4})) which fit the mixing angles. As a result, those values were used to find
the regions for the effective Majorana mass of the electron neutrino, as shown in the Fig. (\ref{f5}).
For this observable, two scattered plots have been only shown since that parameters $m_{1}$ and $a_{\nu}$ are more restrictive for the allowed region.

\subsection{Lepton violation process: $\mu\rightarrow e \gamma$}

In this section, we have calculated the branching ratio for the lepton flavor violation process $\mu\rightarrow e \gamma$ \cite{Akeroyd:2009nu, Lindner:2016bgg} that is mediated by the doubly ($\Delta^{++}$) and singly ($\Delta^{+}$) charged scalars that come from the Higgs triplet (see Eq.(\ref{scal})). The branching ratio \cite{Akeroyd:2009nu} is given by 

\begin{equation}
\textrm{BR}(\mu\rightarrow e \gamma)\approx 4.5\times 10^{-3}\left(\frac{1}{\sqrt{2}v_{\Delta}\lambda}\right)^{4}\left| \left( \mathbf{V}^{\ast}\hat{\mathbf{M}}^{\dagger}_{\nu}\hat{\mathbf{M}}_{\nu} \mathbf{V}^{T}\right)_{e\mu}\right|^{2}\left(\frac{200~GeV}{m_{\Delta^{++}}}\right)^{4}
\end{equation}
where $m_{\Delta^{+}}=m_{\Delta^{++}}\equiv m_{\Delta}$ has been assumed in the previous result. Besides, $\mathbf{V}$ stands for the PMNS mixing matrix.

The branching ratio depends on the PMNS mixing parameters, the single and doubly charged scalars; along with this, the vev of the Higgs triplet takes place. In here, we use the following regions $80~GeV< m_{\Delta}$ and $v_{\Delta}<5~GeV$~\cite{CarcamoHernandez:2018djj}; the PMNS mixing parameters have been already constrained in the previous section, to be more explicit, we use the following regions: $0.01~eV<m_{1}<0.014~eV$, $0.35~MeV<\vert a_{e}\vert<m_{e}$, $0.004~eV<\vert a_{\nu}\vert<0.006~eV$ and $\pi< \eta_{\nu}<6\pi/5$.

\begin{figure}[h!]\centering
	\includegraphics[scale=0.4]{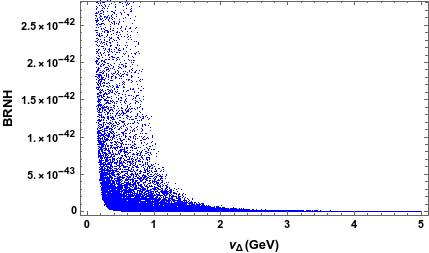}
	\hspace{10mm}\includegraphics[scale=0.4]{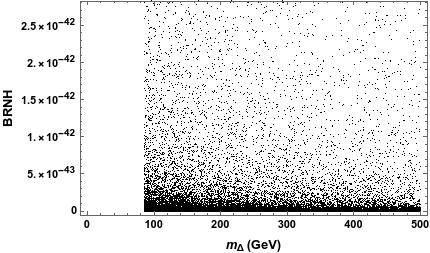}
	\caption{From left to right: BR($\mu\rightarrow e\gamma$) versus the $v_{\Delta}$ and $m_{\Delta}$ parameter. The thick line stands for $3~\sigma$ of C. L.}\label{f6}
\end{figure}

In the Fig. (\ref{f6}), the predicted region is shown
for the branching ratio as function of the vev of the Higgs triplet and the mass of the singly and doubly charged scalars. Our model predicted a region, $\textrm{BR}(\mu \rightarrow e\gamma)\approx 10^{-40}$, that is too much below of the experimental bound $\textrm{BR}(\mu \rightarrow e\gamma)\approx 4.2\times10^{-13}$.

\section{Conclusions}

We have built an economical non-renormalizable lepton model for getting the mixings where the type II see-saw mechanism is responsible to explain tiny neutrino masses. Under a particular benchmark, in the charged lepton sector, the mass matrices have the Fritzsch textures with a shift parameter which makes different to the previous studies.
Our main finding is: a set of values for the relevant parameters was found to be consistent (up to $3~\sigma$) with the last experimental data on
lepton observables for the normal neutrino mass ordering.

To finish, we would like to add that the $\mathbf{S}_{3}\otimes \mathbf{Z}_{2}$ symmetry is an excellent candidate to be the flavor symmetry at low energy. However, one has to look for the best framework where the flavor symmetry solve the majority of open questions on the flavor problem and related issues. In this direction, the quark mixings and the scalar potential analysis will be included to have a complete study but this is a working progress. 

% the model favors the normal hierarchy which is preferred by the current oscillations data. As model prediction, we obtained consistent values for the CP-violating phase and the $\vert m_{ee}\vert$ effective Majorana neutrino mass rate;  the branching ratio for the lepton flavor violation process, $\mu\rightarrow e\gamma$, is well below the current bound.

%\section{Conclusions}
\section*{Acknowledgements}
Garc\'ia-Aguilar appreciates the facilities given by the IPN through the SIP project number 20211170. JCGI thanks  Valentina A. and A. Emiliano  G\'omez Nabor for  sharing great moments and experiences during this long time. This work was partially supported by Project 20211423 and PAPIIT IN109321.

\appendix
\section{$\mathbf{S}_{3}$ flavour symmetry}
The non-Abelian group ${\bf S}_{3}$ is the permutation group of three objects \cite{Ishimori:2010au} and this has three irreducible representations: two 1-dimensional, ${\bf 1}_{S}$ and ${\bf 1}_{A}$, and one 2-dimensional representation, ${\bf 2}$. We list the multiplication rules among them:

\begin{eqnarray}\label{rules}
{\bf 1}_{S}&\otimes& {\bf 1}_{S}={\bf 1}_{S}, \qquad {\bf 1}_{S}\otimes {\bf 1}_{S}={\bf 1}_{S},\qquad {\bf 1}_{S}\otimes {\bf 1}_{A}={\bf 1}_{A}\nonumber\\
{\bf 1}_{A}&\otimes& {\bf 1}_{A}={\bf 1}_{S},\qquad {\bf 1}_{S}\otimes {\bf 2}={\bf 2},\qquad {\bf 1}_{A}\otimes {\bf 2}={\bf 2}\nonumber\\
\begin{pmatrix}
a_{1} \\ 
a_{2}
\end{pmatrix}_{{\bf 2}}
&\otimes&
\begin{pmatrix}
b_{1} \\ 
b_{2}
\end{pmatrix}_{{\bf 2}}=
\left(a_{1}b_{1}+a_{2}b_{2}\right)_{{\bf 1}_{S}} \oplus  \left(a_{1}b_{2}-a_{2}b_{1}\right)_{{\bf 1}_{A}} \oplus	
\begin{pmatrix}
a_{1}b_{2}+a_{2}b_{1} \\ 
a_{1}b_{1}-a_{2}b_{2}
\end{pmatrix}_{{\bf 2}}. 
\end{eqnarray}

\bibliographystyle{bib_style_T1}
\bibliography{references.bib}

\end{document}